\documentclass[aps,prb,twocolumn,groupedaddress,showpacs,showkeys,amssymb]{revtex4}
\usepackage{graphicx}
\def\e{{\rm e}}
\def\i{{\rm i}}
\def\r{\rho}
\def\trace{\mathrm{Tr}}
\newcommand{\lp}{\left(}
\newcommand{\rp}{\right)}
\newcommand{\lac}{\left\{}
\newcommand{\rac}{\right\}}
\newcommand{\lcr}{\left[}
\newcommand{\rcr}{\right]}
\begin{document}
\title{Decoherence in a $N$-qubit solid-state quantum register}
\author{Boris Ischi}
\affiliation{Laboratoire de Physique des Solides, Universit\'e
Paris-Sud, B\^atiment 510, 91405 Orsay, France}
\email{ischi@kalymnos.unige.ch}
\author{Michael Hilke}
\affiliation{Physics Department, McGill University, 3600 rue
University, Montr\'eal, Qu\'ebec, H3A 2T8, Canada}
\email{hilke@physics.mcgill.ca}
\author{Martin Dub\'e}
\affiliation{CIPP, Universit\a'e du Qu\a'ebec \a`a Trois-Rivi\a`eres,
C.P. 500, Trois-Rivi\`eres, Qu\a'ebec, G9A 5H7 Canada}
\email{Martin_Dube@uqtr.ca}
\date{\today}
\begin{abstract}
We investigate the decoherence process for a quantum register
composed of $N$ qubits coupled to an environment. We consider an
environment composed of one common phonon bath and several
electronic baths. This environment is relevant to the
implementation of a charge based solid-state quantum computer. We
explicitly compute the time evolution of all off-diagonal terms of
the register's reduced density matrix. We find that in realistic
configurations, "superdecoherence" and "decoherence free
subspaces" do not exist for an $N$-qubit system. This means that
all off-diagonal terms decay, but not faster than $e^{-q(t) N}$,
where $q(t)$ is of the same order as the decay function of a
single qubit.
\end{abstract}
\pacs{73.21.La,03.67.Lx}
\keywords{Quantum computing, decoherence, N-qubit, coupled quantum
dots}
\maketitle
\section{Introduction}

A typical quantum computer would consist of a large number ($N$)
of two-level quantum systems, coined qubits, where the level
splitting of each qubit and the interaction between pairs of
qubits is adjustable. Quantum operations are then obtained by
varying these parameters along a scheme defined by a quantum
algorithm. The physical system composed by the $N$ two level
systems is our quantum register. In the ideal case, when the
quantum register is isolated, the time evolution of an arbitrary
initial state $\varphi_i$ of the register is unitary. Such an
ideal quantum computer could be used to solve some problems more
efficiently than classical computers. An important example is
Shor's quantum algorithm to factor an integer with $n$ digits in a
time growing polynomially with $n$ instead of exponentially when
using a classical computer.\cite{Shor:1997}

However, any realistic quantum computer is coupled in some way to
an external environment, which leads to decoherence and
dissipation. The quantum register becomes entangled with the
environment, so that its effective evolution is not unitary
anymore. There can also be energy transfers between the register
and the environmental bath, which lead to dissipation and
decoherence. However, dissipation is not a requirement for
decoherence to occur.\cite{Dube:2001}

The quantum decoherence process is elegantly expressed in the
framework of the reduced density matrix of the quantum register.
When no coupling to the environment is present, the reduced density
matrix simply follows a Heisenberg-type evolution. As soon as the
coupling to the environment is introduced, the off-diagonal terms of
the reduced density matrix of the register decay with respect to
time. This is often referred to as phase damping. In the simplest
case of a single two level system connected to an environment, the
off-diagonal elements of the reduced density matrix decay
exponentially in time as $\sim e^{- q(t)}$, where $t$ is the time
and the function $q(t)$ depends on the strength of the coupling to
the environment. In the context of quantum information processing,
such a decoherence event can be expressed as a quantum error.
Following a pioneering work by Shor \cite{Shor:1995}, it was shown
that it is possible to encode each qubit using a minimum of five
qubits (the five qubit code) in conjunction with quantum error
correction algorithm in order to "repair" a faulty
qubit.\cite{Bennett:1996} This would enable an accurate quantum
computation as long as the error rate is small. The drawback of all
quantum error correction schemes is that at least five times as many
qubits are necessary for the same operation than in the ideal case.

In the case where there are $N$ two level systems the situation is
potentially much worst since the decoherence of the register
cannot be simply expressed as a superposition of single qubit
decoherence. Indeed, Palma {\it et al.} argued that the decay of
the most off-diagonal elements goes like $\sim e^{-q(t) N^2}$
(superdecoherence), when all qubits are imbedded in a single
bath.\cite{Ekert:1996} Such a dependence would jeopardize the use
of quantum error correction algorithms as soon as $q(t) N^2$
approaches 1, since the error rate would simply become too large.
A potential rescue to this problem was proposed with the existence
of decoherence free subspaces.\cite{reina:2002}

In this article we investigate the decoherence of $N$ qubits in the
context of a realistic collection of solid state two level systems
(our qubits) imbedded in a semiconducting environment. More
specifically, we consider the case where the two-level system is a
single electron in a double quantum dot patterned in a two
dimensional electron gas confined in a GaAs/AlGaAs heterostructure.
These qubits are coupled to a common phonon bath and to additional
independent electronic baths, representing the metallic leads which
allow the control and operation of the qubits. Experimentally, a
decoherence time of about 1 ns was recently measured in a single
double quantum dot.\cite{Hirayama} For this system, we find that the
decoherence, i.e., the decay of the most off-diagonal elements of
the $N$-qubit reduced density matrix, goes like $e^{-q(t) N}$, a
much slower decoherence rate that previously thought, but that
decoherence free subspaces do not exist.

\begin{figure}[h]
\includegraphics[scale=0.3]{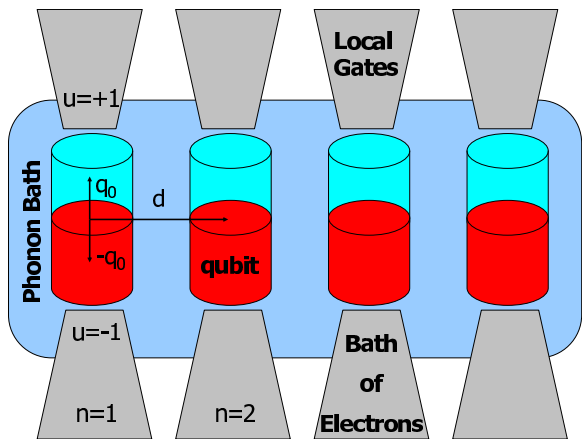}
\caption{Schematic representation of a solid state quantum
computer coupled to a common phonon bath and several independent
electronic baths.} \label{FigureQregister}
\end{figure}

To obtain these results, we consider a model based on a scaled
version of the experimental double quantum dot as seen in Fig.
\ref{FigureQregister}. Tanamoto showed that such a system can
perform all the operations necessary for a quantum computer.
\cite{Tanamoto:2000} Many other groups have also used similar
coupled quantum dots geometries as model system for a
qubit.\cite{Krasheninnikov:1996,Brum:1997,Bandyopadhyay:1998,Zanardi:1998,
Openov:1998,Balandin:1999,Sanders:1999,Openov:1999,Biolatti:2000,Fedichkin:2000}
While we use this particular system for our model, our results are
in fact more general and remain qualitatively similar for different
physical realizations.

The rest of this article is organized as follows. In Section
\ref{SectionHamiltonian} we introduce the Hamiltonian, similar to a
spin-boson model, and describe the $N$-qubit register, the
environment and the coupling between the register and the
environment. In Section \ref{SectionExactFormalExpression} we give
an exact formal expression for the reduced density matrix of the
register. The time evolution of the reduced density matrix is
expressed in terms of the influence functional, which is computed in
Section \ref{SectionInfluenceFunctional}. In these sections the
expressions are fairly general and do not depend on the exact model
considered. In order to gain insight into a physical system we
consider the system shown in Fig. \ref{FigureQregister} in the next
sections. Sections \ref{SectionEquidistantqubits} and
\ref{SectionPhonons} are devoted to the special case of acoustic
phonons coupled to charge-qubits. The coupling to electronic baths
is analyzed in section \ref{SectionCouplinge-e}. In Section
\ref{SectionDecoherenceRateD=0} we evaluate the decoherence rate for
the coupling to a single phonon bath, where we show that the
decoherence function scales as $e^{-q(t)N}$ when increasing the
number, $N$ of qubits. We give physical estimates for piezo and
deformation phonons in Section \ref{SectionPiezoPhonons} and analyze
our results in the dynamical case in Section \ref{SectionDynamics},
where we introduce quantum operations on the register and evaluate
the decoherence process. Finally, a short summary and conclusions
are given in the last section \ref{SectionConclusion}.
\section{The Hamiltonian}\label{SectionHamiltonian}

We use a Hamiltonian for our model qubit, which describes the
tunneling of a single electron tunnelling between two adjacent
quantum dots. The electronic state of the dots can be controlled by
adjusting the gate voltage, which allows either localization of the
electron or resonant tunnelling between the dots. The complete
physical localization of the electron in a given dot is denoted by
the vector $(1,0)^\top$ while localization in the other dot
corresponds to $(0,1)^\top$. At low enough temperatures, only the
combinations of these two states need to be taken into account. Each
qubit is described by the Hilbert space $\mathbb{C}^2$ and the
canonical basis is denoted by $\vert+1\rangle=(1,0)^\top$ and
$\vert-1\rangle=(0,1)^\top$. The single qubit Hamiltonian reads (we
write the Schr\"odinger equation as $\i\dot\varphi=H\varphi$, hence
the units of $H$ are ${\rm s}^{-1}$)
\begin{equation}
H = -\Delta_t \sigma_x - \varepsilon_t \sigma_z\, ,
\end{equation}
where $\sigma_x$ and $\sigma_z$ denote the Pauli's matrices,
$\sigma_z\vert\pm 1\rangle=\pm\vert\pm 1\rangle$ and
$\sigma_x\vert\pm 1\rangle=\vert\mp 1\rangle$. Typically, the
tunnelling matrix element is $\Delta \sim 1-100 GHz$ and the bias
$\varepsilon$ is adjusted with the gate voltage. Both quantities
need to be dynamically controlled for the operation of the qubit
in a quantum computer.\cite{Tanamoto:2000}

For the $N$-qubit register, we write $\vert l\rangle$ (where
$l\in\{-1,+1\}^N$) for the vector of $(\mathbb{C}^2)^{\otimes N}$
defined as $\vert l\rangle:=\vert
l_1\rangle\otimes\cdots\otimes\vert l_N\rangle$. The Hamiltonian
of the register is decomposed as $H^r(t)=\Sigma(t)+\Delta(t)$,
where $\Sigma(t)$ denotes the diagonal part of $H^r(t)$ with
respect to the basis $\vert l\rangle$, that is $\Sigma(t)\vert
l\rangle=\varepsilon(t,l)\vert l\rangle$.

For the total system, {\it i.e.}, the register composed of $N$
qubits plus environment, we consider the following Hamiltonian
\begin{equation}\begin{array}{ll}
H(t)&\displaystyle=H^r(t)+H^e+H^{re}\\[2mm]
H^e&\displaystyle=\sum_{\bf{k}}\omega_k b_{\bf k}^\dag b_{\bf k}\\[4mm]
H^{re}&=\displaystyle \sum_{\bf k}\phi_{\bf k}b_{\bf
k}^\dag+\phi_{\bf k}^\dag b_{\bf k}\, ,
\end{array}\label{EquationTheHamiltonian}\end{equation}
where $\phi_{\bf k}$ are operators acting on the register's
Hilbert space. We assume that they are diagonal with respect to
the basis $\vert l\rangle$. The field operators $b_{\bf k}$ and
$b_{\bf k}^\dag$ are bosonic operators, that is
\begin{equation}
\displaystyle \lcr b_{{\bf k}'},b_{{\bf k}}^\dag\rcr=\delta_{{\bf
k}'{\bf k}}\, ,
\end{equation}
and all other commutators are zero. The Hamiltonian given in Eq.
(\ref{EquationTheHamiltonian}) is very general and is frequently
used to study open quantum systems. It leads directly to the
spin-boson model, which describes many types of environments with
extended degrees of freedom (phonons, electrons, magnons...) and
can be obtained from microscopic models.
\cite{Leggett:1987,Ekert:1996,Weiss:handbook} It is however
inappropriate for localized environments, such as a bath of
nuclear spins. \cite{Stamp:2000}

We decompose $H(t)$ as $H(t)=H_0(t)+\Delta(t)$ with
\begin{equation}\displaystyle H_0(t)=\Sigma(t)+H^e+H^{re}\,
.\end{equation}
Note that $\langle l\vert H_0(t)\vert m\rangle
=\delta_{lm}(\varepsilon(t,m)+H^e+H^{re}(m))$ where
\begin{equation}\begin{array}{ll}
\displaystyle H^{re}(m)&=\displaystyle \sum_{\bf k}\phi_{\bf
k}(m)b_{\bf
k}^\dag+{\phi_{\bf k}(m)}^* b_{\bf k}\\[4mm]
\phi_{\bf k}(m)&\displaystyle=\langle m\vert \phi_{\bf k}\vert
m\rangle\, .\end{array}\label{EquationH^{re}(m)}\end{equation}
As a consequence, we find that the evolution operator associated
to $H_0(s)$ is given by
\begin{equation}
\displaystyle\langle l\vert U_0(t,0)\vert
m\rangle=\delta_{lm}\,\e^{-\i\int_0^t\varepsilon(s,m)ds} \e^{-\i
t(H^e+H^{re}(m))}\, .
\end{equation}

\section{Exact formal expression for the reduced density
matrix}\label{SectionExactFormalExpression}

We now give an exact formal expression for the reduced density
matrix of the register by expanding the evolution operator of the
total system with respect to $\Delta(t)$, the off-diagonal terms
of $H^r(t)$. This corresponds to the method of Leggett {\it et al}
studied at length for the spin-boson model in Ref.
\cite{Leggett:1987}.

The density matrix of the total system is given by
$\r(t)=U(t,0)\r_0 U(t,0)^\dag$ where $\r_0$ is the density matrix
at time $t=0$ and $U(t,0)$ is the evolution operator associated to
the total Hamiltonian $H(s)$. We define the interaction picture
with respect to $H_0(s)$ as $\widetilde{\r}(t)= U_0(t,0)^\dag\r(t)
U_0(t,0)$. Hence, the Heisenberg equation reads
$\i\dot{\widetilde{\r}}(t)=L(t)\widetilde{\r}(t)$, where $L(t)$ is
the Liouville operator $L(t)A=[\widetilde{\Delta}(t),A]$.
Therefore we have $\r(t)=U_0(t,0)({\rm T}\{\exp[-\i\int_0^t L(s)
ds]\}\r_0)U_0(t,0)^\dag$.

The density matrix of the register is defined by tracing out all
the environment degrees of freedom $\r^r(t)=\trace_e[\r(t)]$. So,
we need to compute the trace over the degrees of freedom of the
environment of terms of the form
\begin{equation}
\displaystyle \langle l\vert
U_0(t,0)\widetilde{\Delta}(t^\uparrow_{p-j})\cdots
\widetilde{\Delta}(t^\uparrow_1)\r_0
\widetilde{\Delta}(t^\downarrow_1)\cdots
\widetilde{\Delta}(t^\downarrow_j)U_0(t,0)^\dag\vert m\rangle
,\end{equation}
with $t_1^\uparrow\leq t_2^\uparrow\cdots\leq t^\uparrow_{p-j}$
and $t_1^\downarrow\leq t_2^\downarrow\cdots\leq t^\downarrow_j$.
This term reduces to a sum over the all pair of maps
$(\zeta^\downarrow,\zeta^\uparrow)$ defined on $[0,t]$ with values
in $\{-1,+1\}^N$, constant by step, and with
$\zeta^\uparrow(t^-)=l$ and $\zeta^\downarrow(t^-)=m$. More
precisely, $\zeta^\uparrow$ is constant on each interval
$[t^\uparrow_r,t^\uparrow_{r+1}[$ where $0\leq r\leq p-j$, with
$t_0^\uparrow=0$ and $t^\uparrow_{p-j+1}=t$. Moreover,
$\zeta^\uparrow(t_{p-j}^\uparrow)=l$. On the other hand,
$\zeta^\downarrow$ is constant on each interval
$[t^\downarrow_s,t^\downarrow_{s+1}[$ where $0\leq s\leq j$, with
$t_0^\downarrow=0$ and $t^\downarrow_{j+1}=t$. Moreover,
$\zeta^\downarrow(t_j^\downarrow)=m$. We write
$\zeta^\uparrow_{r}$ for $\zeta^\uparrow(t_{r}^\uparrow)$, and
$\zeta^\downarrow_{s}$ for $\zeta^\downarrow(t_{s}^\downarrow)$.
With these notations, the term above becomes
\begin{equation}\begin{array}{l}
\displaystyle
\sum_{(\zeta^\downarrow,\zeta^\uparrow)}\prod_{r=1}^{p-j}\prod_{
s=1}^j\langle \zeta^\uparrow_r\vert
\Delta(t^\uparrow_r)\vert\zeta^\uparrow_{r-1}\rangle
\langle\zeta^\downarrow_{s-1}\vert
\Delta(t^\downarrow_s)\vert\zeta^\downarrow_s\rangle\\[6mm]
\hspace{1cm}\times\langle\zeta^\uparrow_{p-j}\vert
U_0(t,t^\uparrow_{p-j})\vert\zeta^\uparrow_{p-j}\rangle\cdots
\langle \zeta^\uparrow_0\vert U_0(t^\uparrow_1,0)\vert
\zeta^\uparrow_0\rangle \\[4mm]
\hspace{1cm} \times\langle \zeta^\uparrow_0\vert\r_0\vert
\zeta^\downarrow_0\rangle\langle \zeta^\downarrow_0\vert
U_0(0,t^\downarrow_1)\vert \zeta^\downarrow_0\rangle \cdots
\langle \zeta^\downarrow_j\vert U_0(t^\downarrow_j,t)\vert
\zeta^\downarrow_j\rangle\ .\end{array}\end{equation}

We now assume the register and the environment to be decoupled at
time $t=0$, hence $\r_0=\r_0^r\otimes\r_0^e$. Moreover the
environment is assumed to be initially at thermal equilibrium
which means that $\r_0^e=\e^{-\beta H^e}Z_e^{-1}$, where
$\beta=\hbar/K_B T$, hence $\beta T\simeq7.64\cdot 10^{-12}$ sK.

Let $\gamma$ be the path in the complex plane defined as
$\gamma(s)=s$ for $s\in[0,t]$, $\gamma(s)=2t-s$ for $s\in]t,2t]$,
and $\gamma(s)=-\i(s-2t)$ for $s\in]2t,2t+\beta]$. Define
$\zeta(s)$ as $\zeta(s)=\zeta^\uparrow(\gamma(s))$ for $s\in
[0,t]$, and $\zeta(s)=\zeta^\downarrow(\gamma(s))$ for
$s\in]t,2t]$, and $\zeta(s)=0$ for $s\in]2t,2t+\beta]$. For
$\lambda\geq 0$, we define
\begin{equation}
\displaystyle U_\lambda(s) ={\rm
T}\lac\exp\lcr-\i\int_0^{s}\dot\gamma(s')\lp H^e+\lambda
H^{re}(\zeta(s'))\rp ds'\rcr\rac\, .\label{EquationU_Lambda^b}
\end{equation}
Moreover, we define the influence functional as
\begin{equation}
\displaystyle Z_\lambda[\zeta]={1\over
Z_e}\trace_e[U_\lambda(2t+\beta)]\, .
\end{equation}

Finally, if $p$ and $j$ are integers with $0\leq j\leq p$, we
denote by $\Theta_{pj}$ the set of all pairs $(\Omega_1,\Omega_2)$
of subsets of $\{1,\cdots,p\}$ such that $\Omega_1=\{p_1,\cdots,
p_j\}$ and $\Omega_2=\{q_1,\cdots,q_{p-j}\}$ with
$p_1\leq\cdots\leq p_j$, and $q_1\leq\cdots\leq q_{p-j}$, and such
that $\Omega_1\cup\Omega_2=\{1,\cdots,p\}$ (hence
$\Omega_1\cap\Omega_2$ is empty). Moreover, given $0\leq
t_1\leq\cdots\leq t_p\leq t$ and a pair $(\Omega_1,\Omega_2)$ in
$\Theta_{pj}$, we define the maps
$(\zeta^\downarrow,\zeta^\uparrow)$ as above with
$t_1^\uparrow=t_{q_1},\cdots,t_{p-j}^\uparrow=t_{q_{p-j}}$ and
with $t_1^\downarrow=t_{p_1},\cdots,t_j^\downarrow=t_{p_j}$. We
denote the set of all pairs of maps
$(\zeta^\downarrow,\zeta^\uparrow)$ obtained in that way by
$\Upsilon_{pj}$.

With these definitions, we find the following exact formal
expression for the reduced density matrix of the register
\begin{widetext}\begin{equation}\begin{array}{l}
\displaystyle\langle l\vert\r^r(t)\vert m\rangle=\sum_{p=0}^\infty
(-\i)^p
\int_{0}^t dt_p\int_0^{t_p}dt_{p-1}\cdots\int_0^{t_2}
dt_1\sum_{j=0}^p(-1)^{j}
\sum_{(\zeta^\downarrow,\zeta^\uparrow)\in\Upsilon_{pj}}\\[6mm]
\hspace{1.5cm}\displaystyle
\langle \zeta^\uparrow_0\vert\r^r_0\vert
\zeta^\downarrow_0\rangle\lp\prod_{r=1}^{p-j}\prod_{s=1}^j\langle
\zeta^\uparrow_r\vert
\Delta(t^\uparrow_r)\vert\zeta^\uparrow_{r-1}\rangle
\langle\zeta^\downarrow_{s-1}\vert
\Delta(t^\downarrow_s)\vert\zeta^\downarrow_s\rangle\rp
\exp\lcr-\i\int_0^{t}\lcr\varepsilon(s,\zeta^\uparrow(s))-
\varepsilon(s,\zeta^\downarrow(s))\rcr ds\rcr
Z_1[\zeta]\, .\end{array}\label{EquationExactFormula}
\end{equation}\end{widetext}

Here the elements of the reduced density matrix are expressed in
terms of the influence functional $Z_1[\zeta]$, which describes
the effect of the bath on the time evolution. In the next section
we evaluate this influence functional.
\section{The influence functional}\label{SectionInfluenceFunctional}

To compute the influence functional $Z_1[\zeta]$ we follow the
method used for the spin-boson model in Ref.
\cite{Chakravarty:1985}. We define $g_\lambda(s)=\partial_\lambda
U_\lambda(s)$ and denote by $H_\lambda(s)$ the integrand in Eq.
(\ref{EquationU_Lambda^b}). Then, inverting $\partial_s$ and
$\partial_\lambda$, we obtain
$\i\dot{g_\lambda}(s)=\partial_\lambda \lac H_\lambda(s)
U_\lambda(s)\rac$, hence
\begin{equation}\displaystyle
\i\dot{g_\lambda}(s)=H_\lambda(s)g_\lambda(s)+\dot\gamma(s)
H^{re}(\zeta(s))U_\lambda(s)\, .\end{equation}
Therefore, since $g_\lambda(0)=0$, we have that
\begin{equation}\displaystyle g_\lambda(s)=-\i
U_\lambda(s)\int_0^sU_\lambda^\dag(s')\dot\gamma(s')H^{re}(\zeta(s'))
U_\lambda(s') ds'\, .\end{equation}
As a consequence, defining
\begin{equation}\displaystyle h_\lambda(s)={\i\over Z_e
Z_\lambda[\zeta]}\trace_e\lac U_\lambda(2t+\beta)
U_\lambda^\dag(s)H^{re}(\zeta(s)) U_\lambda(s)\rac ,\end{equation}
we find that
\begin{equation}\displaystyle\partial_\lambda
Z_\lambda[\zeta]=-\lp\int_0^{2t}\dot\gamma(s) h_\lambda(s)ds\rp
Z_\lambda[\zeta]\, ,\end{equation}
hence
\begin{equation}\displaystyle
Z_1[\zeta]=\exp\lcr-\int_0^1 d\lambda\int_0^{2t}\dot\gamma(s)
h_\lambda(s)ds\rcr\, ,
\end{equation}
since by definition, $Z_0[\zeta]=1$.

From Eq. (\ref{EquationH^{re}(m)}) we can decompose $h_\lambda$ as a
sum by defining
\begin{equation}\displaystyle f^+_{{\bf k}}(s)={\i\over Z_e
Z_\lambda[\zeta]}\trace_e\lac U_\lambda(2t+\beta)
U_\lambda^\dag(s) b_{\bf k}^\dag U_\lambda(s)\rac \, ,
\end{equation}
and $f^-_{{\bf k}}(s)$ by the same formula but with the
annihilation operator $b_{\bf k}$ instead of the creation operator
$b_{\bf k}^\dag$. Whence, we have
\begin{equation}\displaystyle h_\lambda(s)=\sum_{\bf k}
\phi_{\bf k}(\zeta(s))f_{\bf k}^+(s) +{\phi_{\bf k}(\zeta(s))}^*
f_{\bf k}^-(s) \, .
\end{equation}

Further, from the Heisenberg equation, defining $\widetilde{b_{\bf
k}^\dag}(s)$ as $U_\lambda^\dag(s) b_{\bf k}^\dag U_\lambda(s)$,
we obtain that
\begin{equation}\displaystyle \i \partial_s \widetilde{b_{\bf
k}^\dag}(s)=\dot\gamma(s)U_\lambda^\dag(s) [b_{\bf k}^\dag,
H^e+\lambda H^{re}(\zeta(s))] U_\lambda(s)\, .\end{equation}
Therefore, from the commutation relations for bosonic operators,
we obtain for $f_{\bf k}^{\pm}$ the following differential
equation
\begin{equation}\displaystyle \partial_s f_{\bf k}^{\pm}(s)=
\pm\dot\gamma(s)\lp\i\omega_k f_{\bf k}^\pm(s)-\lambda\phi_{\bf
k\mp}(\zeta(s))\rp\, ,\label{EquationEqDiffBosons}
\end{equation}
where $\phi_{{\bf k}+}(m)$ is given by Eq.
(\ref{EquationH^{re}(m)}) and $\phi_{{\bf k}-}(m)$ is the complex
conjugate of $\phi_{{\bf k}+}(m)$.

The solution of Eq. (\ref{EquationEqDiffBosons}) is obviously
given by
\begin{equation}\begin{array}{l}
\displaystyle f_{\bf
k}^\pm(s)=\e^{\pm\i\omega_k\gamma(s)}\bigg{(}f_{\bf
k}^\pm(0)+\\[4mm]
\hspace{2.5cm}\displaystyle\left.\mp\lambda\int_0^s
\e^{\mp\i\omega_k\gamma(s')}\dot\gamma(s')\phi_{\bf
k\mp}(\zeta(s'))ds'\rp\,
.\end{array}\label{EquationSolutionEqDiffBosons}
\end{equation}
Now, because of the invariance of the trace under permutations, we
have the boundary condition $f_{\bf k}^\pm(2t+\beta)=f_{\bf
k}^\pm(0)$. Hence, from Eq. (\ref{EquationSolutionEqDiffBosons})
we find that
\begin{equation}\begin{array}{l}
\displaystyle f_{\bf k}^\pm(0)={\lambda
\e^{\pm\beta\omega_k/2}\over
2\sinh(\beta\omega_k/2)}\\[4mm]
\hspace{1.5cm}\displaystyle\times\int_0^t
\e^{\mp\i\omega_ks}\lcr\phi_{{\bf
k}\mp}(\zeta^\uparrow(s))-\phi_{{\bf
k}\mp}(\zeta^\downarrow(s))\rcr ds\, .\end{array}\end{equation}

Defining
\begin{equation}\begin{array}{l}
\displaystyle \Delta_{\bf k}(s)=\phi_{\bf k}(\zeta^\uparrow(s))-
\phi_{\bf k}(\zeta^\downarrow(s))\\[4mm]
\displaystyle \Sigma_{\bf k}(s)=\phi_{\bf k}(\zeta^\uparrow(s))+
\phi_{\bf k}(\zeta^\downarrow(s))\, ,
\end{array}\label{EquationDelta_k(s)=...}\end{equation}
where $\phi_{\bf k}$ is given by Eq. (\ref{EquationH^{re}(m)}), we
find that the influence functional is given by
\begin{equation}
\displaystyle Z^b_1[\zeta]=\e^{-\i X^b}\e^{-\Lambda^b}\,
,\label{EquationZ^b_1[Zeta]}\end{equation}
with
\begin{equation}\begin{array}{l}
\displaystyle\Lambda^b=\sum_{\bf
k}\coth(\beta\omega_k/2)\\
\displaystyle \hspace{1cm}\times{\rm Re}\lcr\int_0^t ds\int_0^s
ds' \e^{-\i\omega_k(s-s')} {\Delta_{\bf k}(s)}^*\Delta_{\bf
k}(s')\rcr \, ,
\end{array}\end{equation}
and
\begin{equation}
\displaystyle X^b=\sum_{\bf k}{\rm Im}\lcr\int_0^t ds\int_0^s ds'
\e^{-\i\omega_k(s-s')}{\Delta_{\bf k}(s)}^*\Sigma_{\bf k}(s')\rcr
\, .
\end{equation}

Recall that $\zeta^\uparrow$ and $\zeta^\downarrow$ are constant
on each interval $[t_{j-1},t_j[$ with $1\leq j\leq p+1$ ($t_0=0$
and $t_{p+1}=t$). We define
\begin{equation}\begin{array}{l}
\displaystyle\Delta_{\bf k}^j=\Delta_{\bf k}(t_{j-1})\\[4mm]
\displaystyle\Sigma_{\bf k}^j=\Sigma_{\bf k}(t_j)\, ,
\end{array}\label{EquationDelta_kSigma_k}\end{equation}
and
\begin{equation}\begin{array}{ll}
q_k(t-t_0)&\displaystyle =\int_{t_0}^t ds\int_{t_0}^s ds'
\e^{-\i\omega_k(s-s')}\\[4mm]
&\displaystyle ={-\i\over\omega_k}(t-t_0)+{1\over\omega_k^2}\lcr
1-\e^{-\i\omega_k(t-t_0)}\rcr\ .\end{array}\end{equation}
Note that
\begin{equation}
\displaystyle \int_{t_{j-1}}^{t_j} ds\int_{t_{m-1}}^{t_m} ds'
\e^{-\i\omega_k(s-s')}=M_k^{jm}\, ,
\end{equation}
where
\begin{equation}\begin{array}{l}
\displaystyle M_k^{jm}=
q_k(t_{j}-t_{m-1})+q_k(t_{j-1}-t_{m})\\[2mm]
\hspace{3cm}\displaystyle-q_k(t_{j}-t_{m})-q_k(t_{j-1}-t_{m-1})\,
.\end{array}\end{equation}
Hence, we find that
\begin{equation}\begin{array}{ll}
\displaystyle\Lambda^b&\displaystyle=\sum_{\bf
k}\coth(\beta\omega_k/2)  \sum_{j=1}^{p+1}
{\rm Re}\lcr q_k(t_j-t_{j-1}){\Delta_{\bf k}^j}^*\Delta_{\bf k}^j\rcr\\[6mm]
&\hspace{3.2cm}\displaystyle+\sum_{j=2}^{p+1} \sum_{m=1}^{j-1}{\rm
Re}\lcr{\Delta_{\bf k}^j}^*\Delta_{\bf k}^m M_k^{jm}\rcr \, ,
\end{array}\label{EquationLambda^b}\end{equation}
and
\begin{equation}\begin{array}{ll}
\displaystyle X^b&\displaystyle=\sum_{\bf k} \sum_{j=1}^{p+1}
{\rm Im}\lcr q_k(t_j-t_{j-1}){\Delta_{\bf k}^j}^*\Sigma_{\bf k}^{j-1}\rcr\\[6mm]
&\hspace{2.5cm}\displaystyle+ \sum_{j=2}^{p+1}\sum_{m=1}^{j-1}{\rm
Im}\lcr{\Delta_{\bf k}^j}^*\Sigma_{\bf k}^{m-1} M_k^{jm}\rcr \, .
\end{array}\label{EquationX^b}\end{equation}

\section{Charge-qubits}\label{SectionEquidistantqubits}

In order to evaluate the influence functional further we need to
specify the form of the coupling in the Hamiltonian
(\ref{EquationTheHamiltonian}). Therefore, we will focus on the
example represented in Fig. \ref{FigureQregister}. We consider $N$
equidistant qubits, with two electronic gates per qubit. We denote
by ${\bf q_n}$ the center of the qubit $n$ ($1\leq n\leq N$, we
put ${\bf q_1}={\bf 0}$). For qubit 1, ${\bf q_{_0}}$ is the
center of the upper dot and ${\bf -q_{_0}}$ the center of the
lower dot ($q_{_0}=\vert {\bf q_{_0}}\vert$). Moreover, we write
${\bf d}={\bf q_2}-{\bf q_1}$, and $d=\vert{\bf d}\vert$, the
distance between qubits. Hence ${\bf q_n}=(n-1){\bf d}$, and ${\bf
q_{_0}}\cdot {\bf d}=0$. Each qubit has a single electron.

For $H^{re}$ we first consider the coupling between electrons in
the qubits and longitudinal phonons. This can be described by a
Fr\"ohlich-type Hamiltonian \cite{Frohlich:54}, where we have
\begin{equation}\displaystyle
\phi_{\bf k}= g(k)\sum_{n=1}^N\e^{-\i {\bf k}\lcr (n-1){\bf
d}+{\bf q_{_0}}\sigma_z^n\rcr}\,
,\label{EquationPhi_k}\end{equation}
where $\sigma_z^n$ is defined as $\sigma_z^n\vert
l\rangle=l_n\vert l\rangle$. This form describes the coupling
$g(k)$ between the charge of qubit $n$ localized at $(n-1){\bf
d}+{\bf q_{_0}}\sigma_z^n$ and the phonon bath. Here, the spatial
configuration is explicitly considered.

Since the optical phonons are gapped at low frequency they do not
contribute to low temperature decoherence, only acoustic phonons are
relevant. We also consider only a linear coupling between the
phonons and the qubits. Nonlinear couplings can also be included and
it has been shown that these can be mapped to an effective
spin-boson model, albeit with decoherence rates (friction
coefficients) that are very strongly temperature dependent
\cite{Dube:1998a}. Surface phonons are not considered either, since
the quantum dots are usually embedded well inside the semiconductor.
Moreover, we assume that the phonons are only coupled to the
diagonal operators of the qubit and we do not consider any
non-diagonal couplings (involving terms like $\sigma_x$ or
$\sigma_y$) between the environment and the qubit. Such terms
appear, e.g., when spin degrees of freedom become relevant
\cite{Dube:2001,Stamp:2000}. For charge qubits, nuclear spins are
irrelevant and we restrict our study to the diagonal spin-boson
model. In Sections \ref{SectionPhonons}, \ref{SectionCouplinge-e},
and \ref{SectionPiezoPhonons}, we will further discuss the acoustic,
electronic and deformation environmental degrees of freedom.

Introducing the notations
\begin{equation}\begin{array}{ll}
\xi(s)&\displaystyle={\lcr\zeta^\uparrow(s)-\zeta^\downarrow(s)\rcr\over
2}\\[2mm]
\chi(s)&\displaystyle={\lcr\zeta^\uparrow(s)+\zeta^\downarrow(s)\rcr\over
2}\, ,
\end{array}\label{EquationDefinitionXiChi}\end{equation}
we obtain that
\begin{equation}
\displaystyle\Delta_{\bf k}(s)=-2\i g(k)\sin({\bf kq_0})
\sum_{n=1}^N\e^{-\i {\bf k}{\bf d}(n-1)}\xi_n(s)\end{equation}
and
\begin{equation}\begin{array}{l}
\displaystyle\Sigma_{\bf k}(s)=2g(k) \sum_{n=1}^N\e^{-\i {\bf
k}{\bf d}(n-1)}\lac \cos({\bf kq_0})\right.\\
\hspace{5cm}\left.\displaystyle-\i\chi_n(s)\sin({\bf kq_0})\rac\,
.
\end{array}\end{equation}
Now, as before, we denote $\xi(t_{j-1})$ as $\xi^j$ and
$\chi(t_m)$ as $\chi^m$. Moreover, we denote by $J$ the $N\times
N$ Jordan bloc $J_{ij}=\delta_{j i+1}$ and write
$\upsilon=(1,\e^{\i{\bf k d}},\cdots,\e^{\i{\bf k d}(N-1)})$. With
these notations, we find that
\begin{widetext}\begin{equation}\begin{array}{ll}
{\Delta_{\bf k}^j}^*\Delta_{\bf k}^m&\displaystyle =4\vert
g(k)\vert^2\sin({\bf kq_0})^2\lp
\langle\xi^j\vert\xi^m\rangle+\sum_{r=1}^{N-1}\langle\xi^j\vert
J^r\vert\xi^m\rangle\e^{-\i {\bf k}{\bf d}r}+\langle\xi^m\vert
J^r\vert\xi^j\rangle\e^{\i {\bf k}{\bf d}r}\rp\, ,
\end{array}\end{equation}
and
\begin{equation}\begin{array}{ll}
{\Delta_{\bf k}^j}^*\Sigma_{\bf k}^{m-1}&\displaystyle =4\vert
g(k)\vert^2\sin({\bf kq_0})^2\lp
\langle\xi^j\vert\chi^{m-1}\rangle+\sum_{r=1}^{N-1}\langle\xi^j\vert
J^r\vert\chi^{m-1}\rangle\e^{-\i {\bf k}{\bf
d}r}+\langle\chi^{m-1}\vert J^r\vert\xi^j\rangle\e^{\i {\bf k}{\bf
d}r}\rp\\[6mm]
&\displaystyle\hspace{7cm}+ {2\vert g(k)\vert^2\sin({\bf
kq_0})\cos({\bf kq_0})\over \sin({\bf k d}/2) }\e^{\i{\bf k
d}/2}\bigg{(} 1-\e^{-\i{\bf k
d}N}\bigg{)}\langle\xi^j\vert\upsilon\rangle
\, .
\end{array}\end{equation}
Hence, introducing the following notations for $r$ between $0$ and
$N-1$ and $n$ between $1$ and $N$
\begin{equation}\begin{array}{ll}
Q^r_{2\pm}(t)&\displaystyle=2\sum_{\bf k}\vert
g(k)\vert^2{1\over\omega_k^2}\coth(\beta\omega_k/2)\sin({\bf
kq_0})^2\lcr\cos({\bf k}{\bf d}r)-\cos(\omega_k t\pm{\bf k}{\bf
d}r)\mp\omega_k t \sin({\bf k}{\bf d}r)\rcr,\\[6mm]
Q_{1\pm}^r(t)&\displaystyle=2\sum_{\bf k}\vert
g(k)\vert^2{1\over\omega_k^2}\sin({\bf kq_0})^2 \lcr \sin(\omega_k
t\pm{\bf k d}r) -\omega_k t\cos({\bf k}{\bf d}r)
 \mp \sin({\bf k}{\bf d}r)\rcr,\\[6mm]
\Psi_n(t)&\displaystyle=2\sum_{\bf k}\vert
g(k)\vert^2{1\over\omega_k^2}\lcr \sin(\omega_k t-{\bf k
d}(n-1/2))-\sin(\omega_k t-{\bf k d}(n-1/2-N))\rcr{\sin({\bf
kq_0})\cos({\bf kq_0})\over \sin({\bf k d}/2) },\\[4mm]
\Phi_n&\displaystyle=2\sum_{\bf k}\vert
g(k)\vert^2{1\over\omega_k}\lcr \cos({\bf k d}(n-1/2))-\cos({\bf k
d}(n-1/2-N))\rcr{\sin({\bf kq_0})\cos({\bf kq_0})\over \sin({\bf k
d}/2) }\, ,
\end{array}\end{equation}
we find that
\begin{equation}\begin{array}{l}
\displaystyle \sum_{\bf k}\coth(\beta\omega_k/2){\rm Re}\lcr
{\Delta_{\bf k}^j}^*\Delta_{\bf k}^m
q_k(t)\rcr=\sum_{r=0}^{N-1}\nu_r\lcr\langle\xi^j\vert
J^r\vert\xi^m\rangle Q_{2+}^r(t)+\langle\xi^m\vert
J^r\vert\xi^j\rangle Q_{2-}^r(t)\rcr\\[6mm]
\displaystyle \sum_{\bf k}{\rm I m}\lcr {\Delta_{\bf
k}^j}^*\Sigma_{\bf k}^{m-1} q_k(t)\rcr=
\sum_{r=0}^{N-1}\nu_r\lcr\langle\xi^j\vert
J^r\vert\chi^{m-1}\rangle Q_{1+}^r(t)+\langle\chi^{m-1}\vert
J^r\vert\xi^j\rangle
Q_{1-}^r(t)\rcr+\sum_{n=1}^N\xi_n^j\lcr\Psi_n(t)-\Psi_n(0)-
\Phi_nt\rcr
\,
,\end{array}\label{EquationSum_kCoth(...}\end{equation}\end{widetext}
where $\nu_r$ is defined as $\nu_0=1$ and $\nu_{r}=2$ for $r\geq
1$. In the next section, we show that in the continuum limit,
$\Psi_n(t)$ and $\Phi_n$ are zero and we provide compact formulas
for $Q^r_{2\pm}(t)$ and $Q_{1\pm}^r(t)$. Note that from Eq.
(\ref{EquationSum_kCoth(...}), the terms $X^b$ and $\Lambda^b$ of
the influence functional (Eqs. (\ref{EquationLambda^b}) and
(\ref{EquationX^b})) can be expressed through the functions
$Q_{1\pm}^r(t)$, $Q_{2\pm}^r(t)$, $\Psi_n(t)$ and $\Phi_n$.
\section{Linear dispersion (acoustic phonons)}\label{SectionPhonons}

In order to simplify the expressions for $Q^r_{2\pm}(t)$ and
$Q_{1\pm}^r(t)$ we need to specify the dispersion relation. Since
we are interested in the effects of a phonon bath, the only
relevant phonons at low temperatures are acoustic phonons, hence
we assume a linear dispersion form $\omega_k\simeq c_L\vert {\bf
k}\vert$, where $c_L$ is the speed of sound in the sample. We
consider a cubic sample of volume $V_S=L_1L_2L_3$. The sum in
$H^e$ and $H^{re}$ runs over all ${\bf k}=2\pi(n_1/L_1,n_2/L_2,$
$n_3/L_3)$ with $n_i$ integers and $\vert {\bf k}\vert< a^{-1}$,
where $a$ is the lattice constant which of the order of a few
angstr\"oms. In the limit of an infinite volume, the sum over
${\bf k}$ can be replaced by and integral
\begin{equation}\displaystyle \sum_{\bf k}\rightarrow{{V_S}\over (2\pi
c_L)^3}\int_0^{c_L a^{-1}} d\omega \omega^2\int_0^\pi
d\theta\sin(\theta)\int_0^{2\pi}d\phi\,
.\label{EquationSum->Integral}\end{equation}
Further, we can assume that $g(\omega/c_L)$ decreases in such a
way that in (\ref{EquationSum->Integral}) we can replace the bound
$c_L a^{-1}$ by $+\infty$.

Recall that ${\bf q_0}$ and ${\bf d}$ are orthogonal. We choose
\begin{equation}\begin{array}{ll}
{\bf k q_0}&\displaystyle ={\omega\over c_L}q_0\cos(\theta)\\[4mm]
{\bf k d}&\displaystyle ={\omega\over
c_L}d\sin(\theta)\cos(\phi)\, .
\end{array}\end{equation}
Since
\begin{equation}\begin{array}{l}
\displaystyle \int_0^{\pi}
\sin(y\cos(\theta))\cos(y\cos(\theta))
g(\sin(\theta))\sin(\theta)d\theta\\[2mm]
\hspace{2cm}\displaystyle=\int_{-1}^{1} \sin(y \tau)\cos(y
\tau)g(\sqrt{1-\tau^2})d\tau=0\, ,
\end{array}\end{equation}
for any function $g$ such that the integral exists, $\Phi_n$ and
$\Psi_n(t)$ are equal to zero.

We now compute the functions $Q_{2\pm}^r$ and $Q_{1\pm}^r$. We
first expand the expressions in $\cos(\omega_kt\pm{\bf kd}r)$ and
$\sin(\omega_k t\pm{\bf kd}r)$ in terms of $\cos(\omega_k
t)\cos({\bf kd}r)$ and $\sin(\omega_k t)\sin({\bf kd}r)$. Further,
because $\int_0^{2\pi} f( \cos(\phi) ) d\phi=0$ for any odd
function $f$, the terms in $\sin({\bf kd}r)$ do not contribute to
the integral. Therefore, we can omit the indices $\pm$. We then
introduce the {\it transit time} ($\tau_s$) and a dimensionless
parameter $\alpha$
\begin{equation}
\displaystyle \tau_s={d\over
c_L}\hspace{.5cm}\text{and}\hspace{.5cm}\alpha={2q_0\over d}\, .
\end{equation}
The parameter $\alpha$ represents the ratio of the size of a qubit
over the distance between qubits (typically, $\alpha$ is smaller
than unity). Integrating over $\phi$ the term in $\cos({\bf kd}r)$
leads to a term $2\pi J_0^B(\omega\tau_s r\sin(\theta))$, where
$J_0^B$ denotes the Bessel function of the first kind. To perform
the integral over $\theta$, we use the formula ($z>0$)
\begin{equation}\begin{array}{l}
\displaystyle \int_0^{\pi}
\sin^2(y\cos(\theta))J_0^B(z\sin(\theta))\sin(\theta)d\theta\\[2mm]
\hspace{3cm}\displaystyle ={\sin(z)\over
z}-{\sin\lp\sqrt{z^2+4y^2}\rp\over\sqrt{z^2+4y^2}}\,
,\end{array}\end{equation}
to find
\begin{equation}\begin{array}{l}
\displaystyle Q_2^r(t)=\int_0^\infty {J_r(\omega)\over
\omega^2}\lcr
1-\cos(\omega t)\rcr\coth(\beta \omega/2)d\omega\\[4mm]
\displaystyle Q_1^r(t)=\int_0^\infty {J_r(\omega)\over
\omega^2}\lcr \sin(\omega t)-\omega t\rcr d\omega
\, ,\end{array}\label{EquationQ1(t)=...Q2(t)=...}\end{equation}
where the spectral function is given by
\begin{equation}
\begin{array}{l}
\displaystyle J_r(\omega)= c_1 \omega^2\vert
g(\omega/c_L)\vert^2\\[4mm]
\hspace{1cm}\displaystyle\times\lcr \frac{\sin(\omega
r/\alpha\omega_q)}{\omega r/\alpha\omega_q}-\frac{\sin(\omega
\sqrt{r^2+\alpha^2}/\alpha\omega_q)}
{\omega\sqrt{r^2+\alpha^2}/\alpha\omega_q}\rcr ,
\end{array}\label{EquationSectralFunction}\end{equation}
with $c_1=V_S/(2\pi^2 c_L^3)$ and
$\omega_q=(\alpha\tau_s)^{-1}=c_L/2q_0$. The spectral function
$J_0(\omega)$ is defined by taking the limit $r\rightarrow 0$ or
\begin{equation}
\displaystyle J_0(\omega)= c_1 \omega^2\vert
g(\omega/c_L)\vert^2\lcr
1-\frac{\sin(\omega/\omega_q)}{\omega/\omega_q}\rcr
.\label{EquationJ_0}\end{equation}
Our expression for $J_0$ has the same form as the one obtained for
a single double quantum dot \cite{Brandes:1999,Fedichkin:2004}.
Moreover, apart from the term linear in $t$ of $Q_1^r(t)$ and the
$r$ dependence, Eq. (\ref{EquationQ1(t)=...Q2(t)=...}) is
identical to Eqs. (4.22a) and (4.22b) in Ref. \cite{Leggett:1987}.
More importantly, the particular $r$-dependence of the spectral
function in Eq. (\ref{EquationSectralFunction}) is the main reason
why the decoherence induced by the bath of phonons does not lead
to ``superdecoherence''. Indeed, if we expand $J_r$ in terms of
$\alpha/r$ we obtain
\begin{equation}J_r(\omega)\simeq -
{c_1\over 2}\lp{\alpha\omega\over r}\rp^2\vert
g(\omega/c_L)\vert^2\cos(r\omega/\alpha\omega_q)\, ,\end{equation}
when $\alpha\omega/(4 r\omega_q)\ll 1$. This $\sim 1/r^2$
dependence is responsible for the suppression of
``superdecoherence''. In the unphysical limit, where $\alpha$, the
ratio of the size of the qubit and the distance between qubits, is
much greater than one, there is no dependence on $r$ for the
expressions of $Q_1^r$ and $Q_2^r$ and the spectral function
$J_r(\omega)$ equals $J_0(\omega)$, which leads to
``superdecoherence''. It is interesting to note that our result
for the influence functional differs from that of references
\cite{Ekert:1996,reina:2002} because they have neglected the angle
between ${\bf k}$ and ${\bf d}$ by introducing the ``transit
time'' as $\omega t_s={\bf k}\cdot{\bf d}$.

The influence functional can now be written in a compact form by
first defining
\begin{equation}
\displaystyle Q^b_{m}(t)=\sum_{r=0}^{N-1}\nu_r\lp
J^r+{J^\dag}^r\rp Q_{m}^r(t)\,
,\label{EquationDefQ^b_2}\end{equation}
with $m=1,2$. Hence, $2Q_m^0(t)$ are the diagonal elements of the
$N\times N$ matrix $Q^b_m(t)$, and $2Q_m^r(t)$, with $r\neq 0$,
the off-diagonal terms. Following a similar notation as in Ref.
\cite{Leggett:1987}, we define
\begin{equation}\begin{array}{l}
\Lambda^b_{jk}= Q^b_{2}(t_{j}-t_{k-1})+Q^b_{2}(t_{j-1}-t_{k})\\[2mm]
\hspace{2.5cm}-Q^b_{2}(t_{j}-t_{k})-Q^b_{2}(t_{j-1}-t_{k-1})\, ,
\end{array}\end{equation}
and $X^b_{jk}$ by the same formula, but with $Q^b_1$ instead of
$Q^b_2$. Note that the part linear in $t$ of $Q_1^b(t)$ does not
contribute to $X^b_{jk}$. With these definitions, we find that
\begin{equation}\begin{array}{l}
\displaystyle\Lambda^b= \sum_{j=1}^{p+1}\langle \xi^j\vert
Q^b_2(t_j-t_{j-1})
\vert\xi^j\rangle+\sum_{j=2}^{p+1}\sum_{k=1}^{j-1}\langle
\xi^j\vert \Lambda^b_{jk}\vert\xi^k\rangle \\[6mm]
\displaystyle X^b=\sum_{j=1}^{p+1}\langle \xi^j\vert
Q^b_1(t_j-t_{j-1})\vert\chi^{_{j-1}}\rangle\hspace{-1.2mm}+
\hspace{-1.2mm}\sum_{j=2}^{p+1}\sum_{k=1}^{j-1}\langle \xi^j \vert
X^b_{jk} \vert\chi^{_{k-1}}\rangle ,
\end{array}
\label{EquationX^b=...Lambda^b=...}
\end{equation}
and the influence functional is then simply given by Eq.
(\ref{EquationZ^b_1[Zeta]}), where $\Lambda^b$ represents the
exponential decay due to the coupling to the bath and $X^b$
describes the phase.

It is important to observe that for a single qubit ({\it i.e.}
$N=1$) we recover the usual formula for the influence functional
of the spin-boson model with the spectral function $J_0(\omega)$.
Expressions (\ref{EquationX^b=...Lambda^b=...}) are evaluated
quantitatively in Sections \ref{SectionDecoherenceRateD=0} and
\ref{SectionPiezoPhonons}.
\section{The coupling to other electrons}\label{SectionCouplinge-e}

The coupling due to the metallic gates can either be provided by
the two dimensional electron gas (lateral gates) or by metallic
top gates. Each gate is considered as a gas of free electrons. The
gates are labelled by $(n,u)$ with $u=+1$ for the gate above the
qubit $n$, and $u=-1$ for the gate below the qubit $n$ (see Figure
\ref{FigureQregister}). The electron gas in the gate $(n,u)$ is
described as
\begin{equation}
\displaystyle H^f_{nu}=\sum_{\bf{k}\,\sigma}E_k {f_{{\bf
k}\sigma}^{nu}}^\dag f_{{\bf k}\sigma}^{nu}\, ,\end{equation}
where $f_{{\bf k}\sigma}^{nu}$ and ${f_{{\bf k}\sigma}^{nu}}^\dag$
are fermionic operators. The gates are supposed to be isolated
from each other, hence fermionic operators with different indices
$n$ or different indices $u$ commute. We describe the coupling
between the register and the electrons in the gates $(n,u)$ as
\begin{equation}
\displaystyle H^{rf}_{nu}=U_{nu}\sum_{{\bf k}{\bf
k'}\sigma}{f_{{\bf k}\sigma}^{nu}}^\dag f_{{\bf k'}\sigma}^{nu}
\end{equation}
where $U_{nu}$ is an operator acting on the register's Hilbert
space
\begin{equation}
\displaystyle U_{nu}= u\sum_{j=1}^NV(\vert j-n\vert)\sigma_z^j\,
.\label{V}
\end{equation}

If we add the sum over $n$ ranging from $1$ to $N$ and $u=\pm 1$
of $H^f_{nu}+H^{rf}_{nu}$ to the Hamiltonian
(\ref{EquationTheHamiltonian}), and compute again the reduced
density matrix of the register, this leads to Eq.
(\ref{EquationExactFormula}), where the bosonic influence
functional is now multiplied by a fermionic influence functional
$Z^f[\zeta]$ which is a product of fermionic influence functionals
corresponding to each gate
\begin{equation}
Z^f[\zeta]=\prod_{n=1}^N\prod_{u=\pm 1}Z^{nu}[\zeta]\, .
\end{equation}

The electron-elec\-tron coupling is given by the Coulomb
potential. Since each gate is much closer to the corresponding
qubit than the distance between qubits, we assume that $V(r)=0$
for $r\geq 1$ in Eq. (\ref{V}). Hence, writing $V_0$ for $V(0)$,
the interaction term reduces to $U_{nu}=uV_0\sigma_z^n$, and we
are left with a two-level system coupled to a fermionic bath by a
contact potential. This model has been studied in Ref.
\cite{Chakravarty:1985}, where it was shown that for a density of
states $\r(\epsilon)$ constant throughout the conduction band
$E_F$ of the bath, {\it i.e.}
$\r(\epsilon)\simeq\r_0\e^{-\epsilon/E_F}$ (hence $\r_0=1/E_F$),
the fermionic bath behaves as a bosonic environment with a
spectral density of the ohmic form. This means, that aside from an
adiabatic shift, {\it i.e.} a shift of the bias energy of the
system, the influence functional $Z^{nu}[\zeta]$ is identical to
the influence functional of the bosonic bath, provided $\phi_{\bf
k}=g_k\sigma_z^n$ with $g_k$ real, and $\sum_{\bf k}
g_k^2\delta(\omega-\omega_k)=J(\omega)$, with a spectral density
function of the ohmic form
\begin{equation}J(\omega)=\eta \omega \e^{-\omega/ \omega_c^f}\,
,\label{EquationJ(omega)Ohmic}
\end{equation}
where
 \begin{equation}
 \eta={2\over \pi^2}\arctan^2(\pi \rho_0 V_0)\, . \label{eta}
 \end{equation}
Here we use $\omega_c^f\simeq  E_F/h$, where $E_F$ is the Fermi
energy of the bath.

As a consequence, the influence functional becomes,
$Z^f[\zeta]=\e^{-\i X^f}\e^{-\Lambda^f}$, using Eqs.
(\ref{EquationDelta_k(s)=...}), (\ref{EquationLambda^b}) and
(\ref{EquationX^b}), where $X^f$ and $\Lambda^f$ are given by Eq.
(\ref{EquationX^b=...Lambda^b=...}) with the matrices
$Q_{_{1,2}}^f(t)=8\;\mathsf{id}\; q_{_{1,2}}^f(t)$, where
$\mathsf{id}$ denotes the identity $N\times N$ matrix, and
$q_{_{1,2}}^f(t)$ is defined as in Eq.
(\ref{EquationQ1(t)=...Q2(t)=...}) with the spectral function given
in Eq. (\ref{EquationJ(omega)Ohmic}). The factor $8$ comes on one
hand from the sum over $u=\pm 1$, and on the other hand from the
factor ${1\over 2}$ in Eq. (\ref{EquationDefinitionXiChi}).
\section{Decoherence function}\label{SectionDecoherenceRateD=0}

We now compute the decreasing rate of the off-diagonal terms of
the register's reduced density matrix $\r^r(t)$. As in Refs.
\cite{Ekert:1996} and \cite{reina:2002}, we consider the case
where the dynamics of the register is trivial, {\it i.e.}
$\Delta(t)\equiv 0$. This means that no quantum operation is
performed. In this case, $\langle l\vert\r^r(t)\vert m\rangle$ is
simply given by the term $p=0$ in Eq. (\ref{EquationExactFormula})
and can be computed exactly. Note that the extension to N qubits
in operation ($\Delta \neq 0$) is far from trivial
\cite{Dube:1998}, and is discussed in Section
\ref{SectionDynamics}. If $\Delta(t)\equiv 0$, we find that
\begin{equation}\begin{array}{l}
\displaystyle \langle l\vert\r^r(t)\vert m\rangle=\langle
l\vert\r_0^r\vert m\rangle \exp\lcr-\i\int_0^t\lcr
\varepsilon(s,l)-\varepsilon(s,m)\rcr
ds\rcr\\[6mm]
\hspace{5cm}\displaystyle \times Z_1^b[l,m]Z^f[l,m]
\, .\end{array}\end{equation}
Moreover, we have
\begin{equation}\begin{array}{ll}
\Lambda^{b}&\displaystyle={1\over 4} \langle l-m\vert Q^b_2(t)
\vert l-m\rangle  \\[2mm]
X^{b}&\displaystyle={1\over 4}\langle l-m)\vert Q^b_1(t)\vert
l+m\rangle \\[2mm]
\Lambda^{f}&\displaystyle=2\Vert l-m\Vert^2
q_2^f(t)\\[2mm]
X^{f}&\displaystyle=0
\, .
\end{array}\end{equation}

For the diagonal terms ({\it i.e.} $l=m$), $\Lambda^b=0$ and
$\Lambda^f=0$, whereas for the off-diagonal terms, with Eq.
(\ref{EquationDefQ^b_2}) we have
\begin{equation}\begin{array}{l}
\displaystyle\Lambda^b={1\over 2}Q_2^0(t)\Vert l-m\Vert^2\\[4mm]
\hspace{1cm}\displaystyle\times\lcr 1+2\sum_{r=1}^{N-1}{\langle
l-m\vert J^r\vert l-m\rangle\over \Vert l-m\Vert^2}{Q_2^r(t)\over
Q_2^0(t)}\rcr
\, .\end{array}\end{equation}
Therefore, for the most off-diagonal terms ({\it i.e.}
$l-m=2(1,\cdots,1)$), we find
\begin{equation}\begin{array}{ll}
\Lambda^b&\displaystyle=2NQ_2^0(t)\lp 1+e(t,N)\rp\\[2mm]
X^b&=0\\[2mm]
\Lambda^f&\displaystyle=8N q_2^f(t)\, ,
\end{array}\label{EquationLambda^b=...(Delta=0)}\end{equation}
with
\begin{equation}\displaystyle e(t,N)=2\sum_{r=1}^{N-1}\lp 1-{r\over N}\rp
{Q_2^r(t)\over Q_2^0(t)}\, ,\label{Equatione(t,N)}\end{equation}
which is small and bounded in the variable $N$ since $Q_2^r\simeq
r^{-2}$ as discussed in section \ref{SectionPhonons}. This implies
that the decreasing rate is proportional to the number $N$ of
qubits in the register. Note that if we neglect the $r$ dependence
and put $Q^r_2(t)=q_2(t)$, for all $r\ge 1$, we find the result
stated in Refs. \cite{Ekert:1996,reina:2002}.

The decoherence functions evaluated above are the ones
corresponding to the most off-diagonal elements of the density
matrix. They correspond to the maximum decoherence rate of the
register. For other matrix elements, the bound $\vert\langle
\xi\vert J^n\vert \xi\rangle\vert\leq\Vert \xi\Vert^2$ leads to
\begin{equation}\displaystyle b_-(t)\leq\vert
\langle l\vert \r^r(t)\vert m\rangle\vert\leq b_+(t)\, ,
\label{EquationBounds}\end{equation}
with
\begin{equation}\begin{array}{l}
\displaystyle b_\pm(t)=\vert \langle l\vert{\r_0^r}\vert
m\rangle\vert\exp\lcr-{1\over 2} Q_2^0(t)\Vert l-m\Vert^2
(1\mp \widetilde{e}(t,N))\rcr\\[6mm]
\displaystyle\hspace{3cm}\times\exp\lcr-2\Vert l-m\Vert^2
q_2^f(t)\rcr
\, ,\end{array}\label{EquationBounds=...}\end{equation}
and
\begin{equation}
\displaystyle\widetilde{e}(t,N)=2\sum_{r=1}^{N-1}{\vert
Q_2^r(t)\vert\over Q_2^0(t)}\, .
\label{Equatione(t,N)Tilde}\end{equation}
Note that $\tilde{e}(t,N)\geq 0$ for all $t$ and $N$. If there is
$\varepsilon<1$ such that $\tilde{e}(t,N)\leq \varepsilon$ for all
$t$ and $N$, the expressions and bounds in Eqs.
(\ref{EquationBounds})-(\ref{Equatione(t,N)Tilde}) imply that there
is no superdecoherence nor a decoherence-free subspace. This result
is consistent with the additivity of decoherence measures in the
short time limit.\cite{Fedichkin:2004b}
\section{Piezo and deformation phonons}\label{SectionPiezoPhonons}

The coupling function $g(k)$ in Eq. (\ref{EquationPhi_k}) depends
on the nature of the phonon coupling. In this section, we discuss
two important cases: the phonon interaction corresponding to the
deformation potential and the piezoelectric phonon interaction.

In experiments on single GaAs/AlGaAs double quantum dots it was
argued that the main contribution to dephasing is due to the
piezoelectric phonon interaction. \cite{Fujisawa:1998} In this
case it was shown that for a double quantum dot the spectral
function is given by \cite{Brandes:1999}
\begin{equation}
\displaystyle J(\omega)=g\omega\lcr 1-{\sin (\omega/\omega_q)
\over \omega/\omega_q}\rcr\e^{-\omega/\omega^b_c}\,
,\label{EquationJ(omega)Piezo}
\end{equation}
where $e^{-\omega/\omega^b_c}$ is the high frequency cut-off
function, which could have different forms. Hence, comparing Eqs.
(\ref{EquationJ(omega)Piezo}) and (\ref{EquationJ_0}) we find that
the piezo case corresponds to taking $c_1\vert
g(\omega/c_L)\vert^2= {g\over\omega} \e^{-\omega /\omega^b_c}$. In
recent experiments on double quantum dots it was found that
$g\simeq 0.03$ \cite{Fujisawa:1998,Brandes:2002,Hirayama}.

For the deformation potential the dependence is given by
\cite{Brandes:1999}
\begin{equation}
\displaystyle J(\omega)=\frac{\omega^3}{\omega_s^2}\lcr 1-{\sin
(\omega/\omega_q) \over
\omega/\omega_q}\rcr\e^{-\omega/\omega_c^b}\,
,\label{EquationJ(omega)deformation}
\end{equation}
where $\omega_s^2\simeq 10^{25}$s$^{-2}$ for typical GaAs values
\cite{Bruus:1993,Fedichkin:2004}. This is exactly the form we
obtain by taking $c_1\vert
g(\omega/c_L)\vert^2=\omega\omega_s^{-2} e^{-\omega/\omega^b_c}$
in Eq. (\ref{EquationJ_0}). In the limit $\omega_q\rightarrow 0$,
both behaviors can be related to the widely used parameterized
form of the spectral function
$J(\omega)\sim\omega^s\e^{-\omega/\omega^b_c}$. Therefore, the
ohmic case $s=1$ is the analogue to the piezoelectric phonon
interaction and the superohmic case $s=3$ is the analogue to the
deformation potential phonon interaction.

\begin{figure}[h]
\includegraphics[scale=0.3]{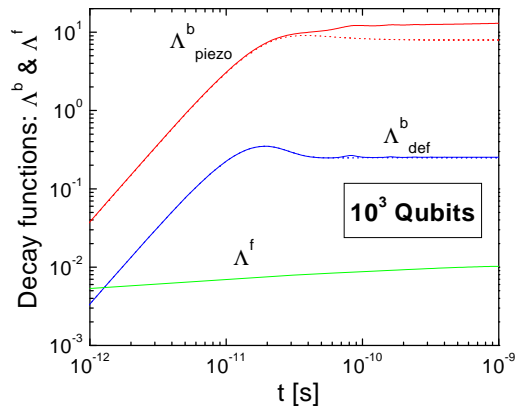}
\caption{Maximum decay functions for $N=10^3$ qubits given by Eqs.
(\ref{EquationLambda^b=...(Delta=0)}) at zero temperature for piezo
and deformation phonons and electronic baths. The constants are
$g=0.03$, $\omega_s^2=10^{25}$s$^{-2}$, $\eta=9.3\cdot 10^{-8}$,
$c_L=5\cdot 10^3$m/s, $q_0$=50nm, $d=400$nm,
$\omega_c^b=\omega_q=5\cdot 10^{10}$s$^{-1}$ (phonons), and
$\omega_c^f=1.3\cdot 10^{15}$s$^{-1}$ (electronic bath). The
corresponding dotted lines are $2NQ_2^{0,piezo}$ and
$2NQ_2^{0,def.}$, which illustrate the expressions for a single
qubit.} \label{FigureLambda}
\end{figure}

We evaluate $\Lambda^b$, $\Lambda^f$ and $Q_2^r$ using Eqs.
(\ref{EquationLambda^b=...(Delta=0)}),
(\ref{EquationQ1(t)=...Q2(t)=...})-(\ref{EquationSectralFunction}),
(\ref{EquationJ(omega)Piezo})-(\ref{EquationJ(omega)deformation}),
and (\ref{EquationJ(omega)Ohmic}) for typical values of a multiple
coupled quantum dots system imbedded in GaAs/AlGaAs for the case
where we have $N=10^3$ qubits. The results are shown in Fig.
\ref{FigureLambda}. The cut-off frequency is important, since it
defines a characteristic time scale. For the electron-phonon
coupling the relevant phonon frequency is given by the smallest
extent of the electronic wave function in the quantum dot, which
we assume to be $\lambda_0$=100 nm. Hence,
$\omega_c^b=c_L/\lambda_0$. This also implies that the coherence
time decreases with a stronger quantum dot confinement. In
contrast, $\omega_c^f$ for the electronic bath coupling is given
by the Fermi energy of the gates. For lateral gates the Fermi
energy is given by the two-dimensional electron system, where we
assume a typical Fermi energy of 8.9 meV, which corresponds to a
Fermi wavelength of 50nm. The coupling constant is given by
(\ref{eta}), where $V_0$ is the charging energy of the dot, which
we assume to be 1.2 meV and corresponds to a typical dot to gate
separation of 100nm. This leads to $\eta=3.2\cdot 10^{-2}$. For a
similar geometry but with top metallic gates, the Fermi energy is
about 5.5 eV (for gold) and the charging energy is similar to the
lateral gates geometry, i.e., 1.2 meV. This leads to
$\eta=9.3\cdot 10^{-8}$. The values of $q_0$ and $d$ are relevant
to recent experiments \cite{Fujisawa:1998}. With these parameters
we obtain a decoherence function for the lateral gates (not shown
in Fig. \ref{FigureLambda}), which is five orders of magnitude
larger than when we consider only top metallic gates (shown as
$\Lambda^f$ in Fig. \ref{FigureLambda}). Hence, we will only
consider the top gate geometry in the remainder of this
discussion.

The main contribution to decoherence is clearly given by the piezo
phonons as was argued earlier \cite{Fujisawa:1998}. The coupling
to the electronic leads (metallic) introduces a smaller
decoherence decay and the form of its time-dependence is the same
as the single qubit case (except for the prefactor). For the
phonon bath, considering $N$ qubits, instead of one qubit,
modifies the form of the time dependence in addition to the
prefactor. The difference in behavior is illustrated by the solid
and dotted lines in Fig. \ref{FigureLambda}.

The small oscillations seen in the figure are reminiscent of
coherence revival \cite{Haroche:1997} and are most likely due to
phase exchange between the qubits via the environmental bath. The
saturation of $\Lambda^b$ at large times is similar to the
saturation seen in the superohmic case of the spin boson
model.\cite{Leggett:1987} This saturation occurs because of the
small density of low frequency modes in the spectral function.
Indeed, at low frequencies, the leading order of the spectral
function for piezo phonons in Eq. (\ref{EquationJ(omega)Piezo}) is
given by $J(\omega)\sim\omega^3$ (superohmic at low frequencies).
Eventually, full decoherence would occur if we include the exchange
of energy between the qubits and the bath ($\Delta\neq 0$). This
introduces another time scale $T_1$ above which all coherence is
lost. However $T_1$ is usually much longer than ${\omega_c^b}^{-1}$,
the typical time-scale of the decoherence due to quantum
fluctuations. We leave the discussion of the energy transfer
processes to Section \ref{SectionDynamics}, where we consider
$\Delta\neq 0$.

\begin{figure}[ht]
\includegraphics[scale=0.3]{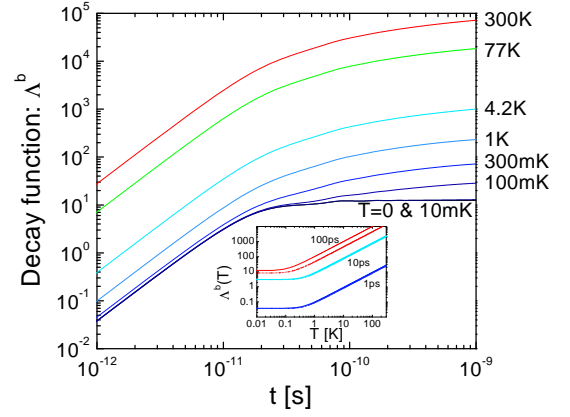}
\caption{Time dependence of $\Lambda^b_{piezo}$ (Eq.
(\ref{EquationLambda^b=...(Delta=0)})) for piezo phonons and
$N=10^3$ qubits for different values of the temperature. Inset:
Temperature dependence of $\Lambda^b_{piezo}$ for 1, 10 and 100ps.
The corresponding dotted lines are the functions
$2NQ_2^{0,piezo}$.} \label{FigureLambdaT}
\end{figure}

The decoherence time due to the quantum and thermal fluctuations
(non-dissipative) can be obtained from Eq.
(\ref{EquationLambda^b=...(Delta=0)}) and is given by
$\Lambda^b(t_{dec})\simeq 1$. Hence, from Fig. \ref{FigureLambda} we
can estimate that $t_{dec}\simeq$ 5ps for $N=10^3$ qubits at zero
temperature. The temperature dependence of the decay function
$\Lambda^b$ (Eq. (\ref{EquationLambda^b=...(Delta=0)})) for the
coupling to the dominant piezo phonons is shown in Fig.
\ref{FigureLambdaT} for $t$=1, 10, and 100ps. As expected, the main
effect of an increasing temperature is to increase the decay
function. At low temperatures the decay function saturates close to
100mK and the decoherence mechanism is only due to quantum
fluctuations. It is interesting to note that for a small number of
qubits, for example $N=10$, the decoherence function at 1ns is only
10\%, which means that there is very little decoherence. However,
this small decoherence would still lead to an error rate during a
quantum operation.\cite{Fedichkin:2004}

%
%

\begin{figure}[h]
\includegraphics[scale=0.34]{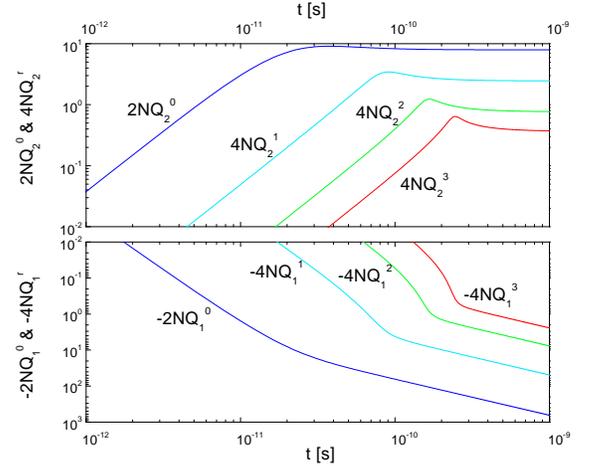}
\caption{The functions $2NQ_1^0$, $2NQ_2^0$, $4NQ_1^r$ and
$4NQ_2^r$ for piezo phonons and $r=1,2,3$ and $N=10^3$.}
\label{Figure4}
\end{figure}

The temperature dependence of the decoherence function for a single
qubit, which is described by the function $Q_2^{0,piezo}$, is shown
in the inset of Fig. \ref{FigureLambdaT} in dotted lines. This shows
that the overall dependence is similar at low temperatures and small
times, where $2NQ_2^{0}$ is actually a good approximation to
$\Lambda^b$. For higher temperatures and longer times the exact
expressions
(\ref{EquationLambda^b=...(Delta=0)})-(\ref{Equatione(t,N)}) have to
be used. In Fig. \ref{Figure4} we plot the functions $Q_1^r$ and
$Q_2^r$ from Eq. (\ref{EquationQ1(t)=...Q2(t)=...}) for the first
values of $r$. Hence, in practice, it is enough to sum over a few
values of $r$ in order to calculate $\Lambda^b$. For the phase
function $X^b$ and $Q_1^r$ the situation is very similar.

%
%

\section{Dynamics in the register}\label{SectionDynamics}

In previous sections \ref{SectionDecoherenceRateD=0} and
\ref{SectionPiezoPhonons} the results were obtained assuming that no
quantum operations are performed ({\it i.e.} $\Delta=0$). In this
section we introduce non-trivial dynamics in order to study the
effects of quantum operations on the decoherence rates.

We assume that $\Delta$ is constant. Suppose that for a set $\Xi$
of qubits, the dynamic is trivial ({\it i.e.} the quantum
operation does not involve those qubits). We take $(\xi,\chi)$ to
be a given path occurring in Eq. (\ref{EquationExactFormula}).
Define $\xi^{tr}$ as $\xi^{tr}_n=0$ if $n$ is not in $\Xi$, and
$\xi^{tr}_n=\xi_n(0)$ otherwise, and $\xi^{dy}$ as $\xi-\xi^{tr}$.
Define $\chi^{tr}$ and $\chi^{dy}$ in the same way. The subscript
$\ ^{tr}$ stands for ``trivial'', and $\ ^{dy}$ for ``dynamical''.
Then, we find that the influence functionals for bosons and
fermions depending only on the trivial part of the path are given
by
\begin{equation}\begin{array}{ll}
\Lambda^{b,f}_{tr}&\displaystyle =\langle\xi^{tr}\vert
Q_2^{b,f}(t)\vert\xi^{tr}\rangle\\[2mm]
X^{b,f}_{tr}&\displaystyle=\langle \xi^{tr}\vert
Q_1^{b,f}(t)\vert\chi^{tr}\rangle\, .
\end{array}\end{equation}
Note that since $Q_1^f(t)$ is a diagonal matrix, $X^f_{tr}=0$.

We now define $\overline{Q}^{b,f}_{1,2}$ as the off-diagonal part
of $Q^{b,f}_{1,2}$. By definition, we have
$\langle\xi^{tr}\vert\xi^{dy}\rangle=0$. As a consequence, we find
that the part of $X^{b}$ depending on cross terms between the
trivial and the dynamical parts of the path is given by
\begin{equation}\begin{array}{l}
\displaystyle x^b=\sum_{j=1}^{p+1}\langle \xi^{dy
j}\vert\overline{Q}_1^{b}(t_{_j})-
\overline{Q}_1^{b}(t_{_{j-1}})\vert\chi^{tr}\rangle\\
\displaystyle\hspace{0.6cm} +\langle\xi^{tr}\vert
\overline{Q}_1^{b}(t-t_{_{j-1}})-
\overline{Q}_1^{b}(t-t_{_j})\vert\chi^{dy j-1}\rangle \, ,
\end{array}\end{equation}
whereas $x^f=0$ since $Q_1^f$ is a diagonal matrix. The
corresponding term for $\Lambda^{b}$, which we denote by
$\lambda^b$, is given by the same formula with $\chi^{tr}$,
$\chi^{dy j-1}$ and $\ _1$ replaced by $\xi^{tr}$, $\xi^{dy j}$,
and $\ _2$ respectively. For the same reason as above,
$\lambda^f=0$. We are interested in the dynamics of the register's
reduced density matrix for time scales smaller than the
decoherence time, which we have estimated in the previous section
($t\leq t_{dec}$). As a consequence, since for $t\leq t_{dec}$,
$Q_1^r(t)$ and $Q_2^r(t)$ are essentially zero for all $r\geq 1$
(see Fig. \ref{Figure4}), we can neglect the cross terms of the
bosonic influence functional, that is we can safely assume that
$x^b=0$ and $\lambda^b=0$.

In general, a quantum computation can be achie\-ved by
succes\-sive single-qubit and C-NOT operations. For C-NOT gates,
we consider the set-up proposed in Ref. \cite{Tanamoto:2000},
described by the Hamiltonian
\begin{equation}
\displaystyle -\Delta{1-\sigma_z\over
2}\otimes\sigma_x-\varepsilon{1+\sigma_z\over 2}\otimes\sigma_z\,
,\end{equation}
where a NOT operation is achieved on the right qubit after a
period given by $\Delta T_{not}={\pi\over 2}$, whenever the left
qubit is in the state $\vert-1\rangle$. Note that in a
semiconductor charge quantum register, only C-NOT operations
between nearest qubits can be achieved. For single-qubit gates, we
consider the Hamiltonian $-\Delta\sigma_x-\varepsilon\sigma_z$.
Three single-qubit operations are needed for a quantum
computation. Two of them are trivial, {\it i.e.} $\Delta=0$ and
$T_1\varepsilon=3\pi/4$ and $T_2\varepsilon=7\pi/8$. The last one
is the Hadamar gate given by $\Delta=\varepsilon$ and
$T_{Had}\sqrt{\Delta^2+\varepsilon^2}={\pi\over 2}$.

As a consequence, the decay rate of the most off-diagonal terms of
the register's density matrix for a single C-NOT operation or
qubit rotation, will remain proportional to $N$. For parallel
qubit rotations or C-NOT operations, the compact formulas for the
bosonic influence functional are a good starting point to discuss
a generalized NIBA approximation. However, a more detailed
analysis is beyond the scope of this article.
\section{Conclusion}\label{SectionConclusion}

We have analyzed the decoherence process in a solid state quantum
register with $N$ qubits. We showed that the decay rate of the
most off-diagonal terms of the register's density matrix is
proportional to $N$ in all situations relevant to a scaled charge
solid-state quantum computer, where the qubits are coupled to a
common phonon bath and to independent electronic gates.  We
obtained compact expressions for the $N$-qubit decoherence
functions and argued that when performing quantum operations the
decoherence function follows a very similar dependence as compared
to the static case.

B.I. acknowledges support from the Swiss National Science
Foundation and M.H. support from NSERC, FCAR and RQMP.
\bibliographystyle{apsrev}
\bibliography{IHD}
\end{document}